# Developing a Suitability Assessment Criteria for Software Developers: Behavioral Assessment Using Psychometric Test

JAYATI GULATI, Guru Gobind Singh Indraprastha University, School of Information and Communication Technology, India

BHARTI SURI, Guru Gobind Singh Indraprastha University, School of Information and Communication Technology, India

LUIZ FERNANDO CAPRETZ, Western University, Department of Electrical and Computer Engineering - Software Engineering, Canada

BIMLESH WADHWA, School of Computing, National University of Singapore, Singapore

ANU SINGH LATHER, Guru Gobind Singh Indraprastha University, School of Information and Communication Technology, India

A suitability assessment instrument for software developers was created using a psychometric criteria that identify the impact of behavior on the performance of software engineers. The instrument uses a questionnaire to help both individuals and IT recruiters to identify the psychological factors that affect the working performance of software engineers. Our study identifies the relationship between the behavioral drivers and the programming abilities of the subjects. In order to evaluate the instrument, a total of 100 respondents were compared on the basis of their programming skills and nine behavioral drivers. It was concluded that there is a direct relationship between certain human qualities, such as "Attention to Detail," and the programming style of the students, while the "Locus of Control" factor was observed to have a negative correlation with performance in programming.

## 1 INTRODUCTION

The software development process involves human beings at each stage. This necessitates a thorough study of personality traits in the software industry. The study of the human psyche in software engineering is an interdisciplinary research field focusing on human psychology and its impact on software and its development process. According to psychological research, emotions and mood deeply influence the cognitive abilities and performance of workers, including creativity and analytical problem solving [5].

Though the impact of human behavior on the software development process is significant, this factor has been neglected by researchers and professionals in the field of software engineering, only in the last 10 years the topic has been receiving increased attention. Due to the overlook of these factors, the quality of the process may be lowered, thus affecting the end product [7] [11]. The emerging relevance to carry out this kind of study was based on the importance to determine whether behavior has a significant impact on the working style of a software engineer. This raises the need to identify and categorize behavioral drivers and their impact on the efficiency of software developers as well as the development process. Hence, the study of factors like human intellect, skills, patience, discipline, etc. is important as they may have a significant effect on the quality of the process and the final product. The important role of these human qualities suggest the need for certain desirable behavioral drives in software developers and the importance of evaluating the capability of software developers. The work considers nine personality traits: patience (P), teamwork (T), attention to detail (AD), responsibility and ownership (RO), locus of control (LC), communication skills (CS), commitment and perseverance (CP), openness to change (OC), and a do-it-now approach (DIN). These factors are described in table 1. A questionnaire consisting of 100 questions based on these factors is used for the study, with questions corresponding to the nine behavioral drivers. Appendix discusses the questions.

Software engineering is a discipline dedicated to develop and maintain high quality software. It lays down standards, procedures, best practices, and models to develop high quality software. The development process involves human beings at every stage of the software development life cycle (SDLC) and, therefore, human presence is inevitable in software development. During the process, individuals are assigned different tasks based on their domain knowledge and capabilities. Several researchers have studied the impact of personality [1], [3], [2] emotional intelligence [6], attitude and behavior in the software development process [4], [8]. The most common instruments used in software psychology are the Myers-Briggs Type Indicator [9] and the Big-Five model [10]. The intent of this work is to assess if the human attributes listed in table 1 affect the programming capabilities of individual or not.

Table 1. Behavioral Components used in the model and their description.

| Behavioral Components | Description |
|---|---|
| Patience (P) | It is the state of enduring under different circumstances without showing annoyance/anger in a negative way. |
| Communication Skills (CS) | They are the skills required to pass information through the exchange of ideas, perceptions, and commands. |
| Teamwork (T) | It is defined as the ability to work or interact in groups together for their common/mutual benefit, as opposed to working in competition for selfish benefit. |
| Do-it-now Approach (DIN) | This factor defines the eagerness and enthusiasm of a person when new challenges come up. |
| Responsibility and Ownership (RO) | Responsibility is the ease with which a person takes lead and shoulders the workload. Ownership is defined by the ability to own mistakes, accept them, and work towards their improvement. |
| Commitment and Perseverance (CP) | Commitment is about keeping up with promises and agreements. Perseverance is sticking to something, independently of the time needed to complete the activity or any unfavorable situation. |
| Attention to Detail (AD) | This aspect of human nature deals with completion of the given task while paying extra attention to minute details. |
| Openness to Change (OC) | This quality ensures that a person accepts changes to improve the task, without being egoistic about their own work. |
| Locus of Control (LC) | This factor depicts human perception about the events in their life and the extent to which they believe they can control them. |

## 2 RESEARCH METHODOLOGY

The methodology is based on gathering data from software engineering students using a survey (refer to the questionnaire in the Appendix). Additionally, their respective teachers from the university were required to assess students on their programming skills based on their performance in programming lab assignments. The students were rated on a scale of 1-5 based on the correctness of their programming logic and coding speed, where '5' corresponds to highest mark and '1' indicates the lowest mark.

### 2.1 Objectives
- To study the relationship between the student's programming skills and the behavioral criteria, i.e., P, T, AD, RO, CS, CP, OC, and DIN.
- To study the relationship between a student's performance and LC.

### 2.2 Hypothesis
- There is a positive correlation between the student's performance and the behavioral drivers, i.e., P, T, AD, RO, CS, CP, OC, and DIN.
- There is a negative correlation between the student's performance and LC.

### 2.3 Data Collection

In the present study, software engineering students were surveyed for personality trait assessment. The survey comprised 100 questions based on the findings of researchers and psychologists representing a particular personality trait. A total score was generated by answering the complete questionnaire. The survey was conducted online using the Talent Power tool[1] in a university environment. The complete data corresponding to all answered questions was collected, filtered and cleaned. The final sample size was 100 software engineering students. The students were rated on a scale of 1 to 5 in terms of finding an optimized solution for a given programming problem. The scores from the survey and those given by their respective teachers were compared. SPSS statistical tool was used for analyzing the data and assessing the correlation between the personality types and programming skills of software engineering students. 77% of the subjects were males and 23% were female. Age was taken as a categorical variable represented by (1): 18-20; (2): 21-22; (3): 22+. Schooling was also taken as a categorical value represented by (1): convent; (2): government or public; (3): private.

## 3 RESULTS

The data gathered from the survey of 100 software engineering students was analyzed using SPSS Data Editor Tool. A clustering algorithm was used to form groups among a given data set based on certain fixed characteristics. The main idea was to define one "k" center for each cluster. Through a fixed number of iterations, the data set aligns to a respective center point belonging to a cluster. We use k-means clustering algorithm defining k=3, because this value of k gives the best possible results, i.e., the entire data set is divided into non-overlapping values. After nine iterations to cluster the data, it was empirically seen that three clusters gave sparse values of nine personality traits as compared to other clusters, hence the number of clusters was set to three. It can be seen that the final clusters are represented by the highest teacher ratings with P having the highest value, followed by RO, CS, DIN, AD, T, OC, CP, and LC. Likewise, second cluster number corresponds to low ratings in programming lab assignments. Subsequently, regression analysis was performed. The best possible values of characteristics P, T, AD, RO, LC, CS, CP, OC, DIN were predicted. The corresponding values of P, T, AD, RO, LC, CS, CP, OC, and DIN, in that order, were predicted and are displayed in table 2. According to the regression equation used, the values listed above of personality characteristics would give the best performance in programming.

---
[1] Talent Power, 2017, available at http://www.talentmanpower.com/ques/main.html

Table 2. Predicted values

| P | 4.89 | RO | 4.23 | CS | 4.10 |
|---|---|---|---|---|---|
| AD | 3.835 | CP | 3.545 | OC | 3.861 |
| DIN | 3.931 | T | 3.651 | LC | 2.889 |

## 4 CONCLUSIONS

The results demonstrated that P, RO, and CS were the behavioral drivers that most directly affected the efficiency of the developer. They were followed by, in this order, DIN, AD, T, OC, and CP. The next significant factors that had an impact on the competency of a developer were the DIN, OC, and AD in this order. LC was found to have a negative correlation with teacher ratings. It was observed that highly-rated students did not score high on LC. The findings were in line with the initial hypotheses. Therefore, regression analysis helped in predicting the values of dependent variables given the values of independent variables. The prediction shows that P was observed to be the most important factor for achieving the best skill set in programming. Applying the clustering technique showed that a majority of students were average in their programming skills. Teacher ratings form the cluster centers; three clusters were formed with high, average, and low teacher ratings. In this investigation P was depicted as the most important behavioral driver required for being a competent software developer. And LC was seen to be least related to the performance of a software developer. Hence, it can be concluded from the study that four of the factors are correlated to programming practices, when practiced by an individual, while one of the factors has a negative correlation with programming skills. This work may benefit education, practice, and research in software psychology. The study highlights individual qualities, which might help to improve programming skills and brushing up on some of the qualities that directly impact on their grades. Additionally, the software industry could gain insights about hiring employees through this study. Finally, researchers could use the same study for deeper investigations aimed at finding other factors from different psychological tests which may help improve the overall quality of the software.

## 5 APPENDIX

Sample questions that assess the nine drivers of the model are preseneted as follows with multiple choice answers using Likert scale.

I. Patience, 2 out of 6 questions.

Q.1. If you had to share a room with a distant cousin for a week:

a) You hesitate 3

b) You Refuse 1

c) You agree immediately 5

Q.2. Your friend arrives 45 minutes late for an appointment:

II. Good Communication Skills, 3 out of 15 questions.

Q.1. I show genuine interest when people are talking to me, whatever the subject or topic may be.

Q.2. I look at the feeling behind the words people are using.

Q.3. I avoid judging the other person while he is speaking.

III. Cooperation with Peers, 3 out of 16 questions. Q.1. I participate in teams but avoid them when I can.

Q.2. When working in a team, I prefer to take up individual assignments.

Q.3. I prefer shorter meetings and sometimes find myself drained after meetings.

IV. Do It Now Approach, 2 out of 7 questions.

Q.1. I love starting new projects, especially "Impossible" ones?

Q.2. I quickly lose interest in a project or job once it is up and running?

V. Responsibility and Ownership, 2 out of 9 questions.

Q.1. I see myself as someone who does a thorough job.

Q.2. I see myself as someone who can be somewhat careless.

VI. Commitment and Perseverance, 3 out of 14 questions.

Q.1. Regardless of whether I work for myself or someone else, there is no change in my level of efforts.

Q.2. I do not compromise on the quality of whatever work I do.

Q.3. I usually find myself cramming my lessons.

VII. Attention to Detail, 3 out of 11 questions.

Q.1. I can describe myself as a person who goes into every details of a project.

Q.2. I can describe myself as person who is short and precise.

Q.3. I just do not notice the little things that other people do.

VIII. Openness to Change, 2 out of 5 questions.

Q.1. I see myself as someone who is original, comes up with new ideas.

Q.2. I see myself as someone who is curious about many different things.

IX. Locus of Control, 2 out of 12 questions.

Q.1 Whether or not I get to be a leader depends on my ability.

Q.2. My life is controlled by accidental happenings.